\documentclass[11pt]{article}
\pdfoutput=1
\newcommand{\Comment}[1]{{}}
\usepackage{amsfonts,amsthm,amsmath,amssymb,slashed}
\usepackage[textwidth = 430 pt, textheight = 630 pt]{geometry}
\usepackage{color}
\usepackage{booktabs}
\usepackage{subcaption}
\usepackage[export]{adjustbox}

\Comment{\usepackage{color}
\definecolor{MyDarkBlue}{rgb}{0.15,0.15,0.45}
\usepackage[linktocpage=true]{hyperref}
\hypersetup{
colorlinks=true,
citecolor=MyDarkBlue,
linkcolor=MyDarkBlue,
urlcolor=MyDarkBlue,
pdfauthor={Horatiu Nastase},
pdftitle={Coupling ModMax theory precursor with scalars and BIon-type solutions},
pdfsubject={hep-th}
}
\usepackage[utf8]{inputenc}
\usepackage[numbers,sort&compress]{natbib}
\usepackage{hypernat}}
\usepackage{graphicx}
\usepackage{cite}
\usepackage[colorlinks=true]{hyperref}
\hypersetup{
  allcolors = blue,
}

\newcommand\ignore[1]{}
\def\one{{\,\hbox{1\kern-.8mm l}}}

\def\a{\alpha}\def\b{\beta}

\def\d{\partial}

\newcommand{\Cset}{{\,\,{{{^{_{\pmb{\mid}}}}\kern-.45em{\mathrm C}}}}}

\newcommand{\be}{\begin{equation}}
\newcommand{\bea}{\begin{eqnarray}}

\newcommand{\ee}{\end{equation}}
\newcommand{\eea}{\end{eqnarray}}

\begin{document}

\renewcommand{\thefootnote}{\fnsymbol{footnote}}

\makeatletter
\@addtoreset{equation}{section}
\makeatother
\renewcommand{\theequation}{\thesection.\arabic{equation}}

\rightline{}
\rightline{}




\begin{center}
{\LARGE \bf{\sc Coupling ModMax theory precursor with scalars, and BIon-type solutions}}
\end{center}
 \vspace{1truecm}
\thispagestyle{empty} \centerline{
{\large \bf {\sc Horatiu Nastase${}^{a}$}}\footnote{E-mail address: \Comment{\href{mailto:horatiu.nastase@unesp.br}}{\tt horatiu.nastase@unesp.br}}
                                                        }

\vspace{.5cm}

\centerline{{\it ${}^a$Instituto de F\'{i}sica Te\'{o}rica, UNESP-Universidade Estadual Paulista}}
\centerline{{\it R. Dr. Bento T. Ferraz 271, Bl. II, Sao Paulo 01140-070, SP, Brazil}}

\vspace{1truecm}

\thispagestyle{empty}

\centerline{\sc Abstract}

\vspace{.4truecm}

\begin{center}
\begin{minipage}[c]{380pt}
{\noindent Recently, the most general theory of electromagnetism invariant under duality and conformal invariance was written, and 
dubbed ModMax. It arises from a generalization of Born-Infeld (BI) theory by taking the infinite tension limit, $T\rightarrow\infty$. 
In this note we show that this generalization can be obtained from a brane-like construction, just like BI, and can thus be coupled to 
scalars in the same way to obtain a DBI-like action. 
All the BIon and catenoid solutions, and their interpolations, are still solutions of the generalized DBI-like action, suggesting that an 
interpretation within string theory could be possible. We also show that Ra\~nada's knotted solutions (with nonzero helicities)
are still valid, both in the ModMax theory, and in its precursor.
}
\end{minipage}
\end{center}

\vspace{.5cm}

\setcounter{page}{0}
\setcounter{tocdepth}{2}

\newpage

\renewcommand{\thefootnote}{\arabic{footnote}}
\setcounter{footnote}{0}

\linespread{1.1}
\parskip 4pt



\section{Introduction}

In a remarkable paper \cite{Bandos:2020jsw}, the unique conformally invariant and duality invariant extension of Maxwell electromagnetism 
was found (both in Lagrangian formalism for ${\cal L}(\vec{E},\vec{B})$, and in Hamiltonian formalism for ${\cal H}(\vec{D},\vec{B})$), 
depending on a dimensionless (numerical) parameter $\gamma$, reducing to Maxwell's theory at $\gamma=0$.\footnote{For an
alternative derivation of this result see \cite{Kosyakov:2020wxv}.} The Hamiltonian was 
found as a $T\rightarrow\infty$ limit of a generalization of the Born-Infeld theory \cite{Born:1934gh}
with the same parameter $\gamma$ (such that 
$T\rightarrow\infty$ gives the Maxwell theory at $\gamma=0$). $P$-form generalizations were given in \cite{Bandos:2020hgy}, and a 
supersymmetrization was presented in \cite{Bandos:2021rqy} (see also \cite{Kuzenko:2021cvx}). 
A number of papers studied the solutions and properties of the ModMax 
theory (coupled to gravity), among them \cite{Flores-Alfonso:2020euz,BallonBordo:2020jtw,Flores-Alfonso:2020nnd,Amirabi:2020mzv,Babaei-Aghbolagh:2020kjg,Neves:2021tbt,Bokulic:2021dtz,Cano:2021tfs,Dassy:2021ulu,An:2021plu,Kruglov:2021bhs,Zhang:2021qga}.

But it is known that the Born-Infeld theory can be coupled to scalars, and the resulting Dirac-Born-Infeld (DBI) theory has the simple 
interpretation in string theory as the theory for the massless modes appearing on a D-brane. As such, a supersymmetrization is 
available, and in terms of the scalars representing brane fluctuations and the gauge fields, one has both an extremal (supersymmetric)
solution, the BPS BIon \cite{Callan:1997kz}, together with the original BIon of Born and Infeld (devised as a non-singular alternative 
to the Dirac electron), and a "catenoid" solution. Both the BPS BIon and catenoid have interpretations in terms of brane geometries, 
and there are interpolating solutions as well \cite{Gibbons:1997xz,Gibbons:2001gy,Nastase:2005pb}.

A relevant question that we are trying to answer here is then: can one interpret the precursor of ModMax, the generalization of the 
Born-Infeld action, also as a brane-type action? And are the same kind of solutions still valid, and if so, are they modified? 
To the first question we will find a positive answer, with a certain caveat, whereas to the second question we will find a positive answer, 
with no modification for the solution. 

We will also consider whether knotted solutions of electromagnetism found by Ra\~nada \cite{Ranada:1989wc,ranada1990knotted}
are still solutions of ModMax theory and its precursor, and we will find that they are.

The paper is organized as follows.
In section 2 we will describe the BI generalization and the ModMax limit, in section 3 we will re-write it in terms of a brane-like 
description, which will allow us to show that the same solutions are still valid, including the Ra\~nada solutions, 
and in section 4 we will conclude.

\section{Born-Infeld duality-invariant generalization and ModMax electromagnetism}

The generalization of the Born-Infeld theory has the Hamiltonian
\be
{\cal H}(T,\gamma;\vec{D},\vec{B})=\sqrt{T^2+2T\left(s\cosh \gamma-\sinh \gamma \sqrt{s^2-p^2}\right)+p^2}-T\;,\label{genHam}
\ee
where $T$ is a parameter of dimension 4, $\vec{D}$ is the electric induction, and
\be
s=\frac{\vec{D}^2+\vec{B}^2}{2}\;,\;\;
p=|\vec{D}\times\vec{B}|=\sqrt{\vec{D}^2\vec{B}^2-(\vec{D}\cdot\vec{B})^2}.
\ee

The Lagrangian is, as usual, the Legendre transform of the Hamiltonian with respect to $\vec{D}$, 
\be
{\cal L}(T,\gamma;\vec{E},\vec{B})=\vec{E}\cdot\vec{D}-{\cal H}(T,\gamma;\vec{D},\vec{B})\;,
\ee
though an explicit formula was not given in \cite{Bandos:2020jsw}. As usual, the Legendre transform implies the constitutive relations
\bea
\vec{E}=\frac{\d {\cal H}}{\d \vec{D}}\;, &&\vec{H}=\frac{\d{\cal H}}{\d \vec{B}}\cr
\vec{D}=\frac{\d {\cal L}}{\d \vec{E}}\;, && \vec{H}=-\frac{\d {\cal L}}{\d \vec{B}}\;,
\eea
and, in terms of $\vec{E},\vec{D},\vec{B},\vec{H}$, the equations of motion take the form of the equations 
of motion inside a material (in the presence of sources, in 
$\vec{\nabla}\cdot\vec{D}=\rho_{\rm ext}/\epsilon_0$ we obtain only the {\em external} charge,
delta function sources, introduced as an extra term in the Lagrangian, 
whereas in $\vec{\nabla}\cdot \vec{E}=\rho/\epsilon_0$ we also obtain charges due to the polarization 
of the material, or in this case, of the vacuum, so the charge density is spread out) 
\bea
\vec{\nabla}\times \vec{E}=-\frac{1}{c}\d_t \vec{B}\;, && \vec{\nabla}\cdot \vec{B}=0\;,\cr
\vec{\nabla}\times \vec{H}=\frac{1}{c}\d_t \vec{D}\;, && \vec{\nabla}\cdot \vec{D}=0.\label{Mxmedium}
\eea

In \cite{Bandos:2020jsw}, it is also shown that  conditions for conformal invariance are 
\be
\vec{E}\times \vec{H}=\vec{D}\times \vec{B}\;,\;\;
\vec{D}\cdot\vec{E}+\vec{B}\cdot \vec{H}=2{\cal H}\;,
\ee
while a condition for duality is
\be
\vec{E}\cdot\vec{B}=\vec{D}\cdot\vec{H}.
\ee

In the Appendix we quickly review the case of the BI action, that we will be generalizing here.

\subsection{ModMax theory}

The  conformal and duality invariant ModMax theory is obtained from the low energy, or $T\rightarrow \infty$, limit of the 
general theory (\ref{genHam}), so 
\be
{\cal H}(\gamma; \vec{D},\vec{B})=s\cosh \gamma-\sinh \gamma\sqrt{s^2-p^2}.
\ee

In this case also the Legendre transform can be made explicitly, as was done in \cite{Bandos:2020jsw}, and the result is 
the Lorentz invariant Lagrangian
\be
{\cal L}^{\rm ModMax}(\gamma;\vec{E},\vec{B})=\cosh\gamma S+\sinh \gamma\sqrt{S^2+P^2}\;,
\ee
where we have defined the quantities $S$ and $P$ as in  \cite{Bandos:2020jsw},
\bea
S&=&\frac{\vec{E}^2-\vec{B}^2}{2}=-\frac{1}{4}F_{\mu\nu}F^{\mu\nu}=-\frac{b^2F}{2}\cr
P&=& \vec{E}\cdot\vec{B}=-\frac{1}{4}F_{\mu\nu}\tilde F^{\mu\nu}=b^2G\;,
\eea
and we have also related to the quantities $F$ and $G$ defined for the BI electromagnetism (see the Appendix).

\subsection{Lagrangian for the general theory}

For the general theory (\ref{genHam}), we calculate
\be
\vec{E}=\frac{\d {\cal H}}{\d \vec{D}}=\frac{T\left[\cosh\gamma \vec{D}-\frac{\sinh \gamma}{\sqrt{s^2-p^2}}\left(s\vec{D}-(\vec{D}\vec{B}^2
-\vec{B}(\vec{B}\cdot\vec{D}))\right)\right]+\vec{D}\vec{B}^2-\vec{B}(\vec{B}\cdot\vec{D})}{\sqrt{T^2+2T\left(\cosh \gamma s-\sinh \gamma 
\sqrt{s^2-p^2}\right)+p^2}}\;,
\ee
and then
\bea
{\cal L}(T,\gamma;\vec{E},\vec{B})&=&\vec{D}\cdot\vec{E}-{\cal H}(T,\gamma;\vec{E},\vec{B})\cr
&=&T+\frac{-T^2+T\left[-\cosh\gamma \vec{B}^2-\sinh\gamma \frac{s\vec{B}^2+p^2}{\sqrt{s^2-p^2}}\right]}{
\sqrt{T^2+2T\left(\cosh \gamma s-\sinh \gamma \sqrt{s^2-p^2}\right)+p^2}}\;,
\eea
but this has to be rewritten in terms of $\vec{E},\vec{B}$. 

A natural guess for that is 
\bea
{\cal L}&=&T-\sqrt{T^2-2T\left(S\cosh\gamma+\sinh\gamma \sqrt{S^2+P^2}\right)-P^2}\cr
&=&T\left[1-\sqrt{1+F\cosh\gamma -\sinh\gamma\sqrt{F^2+4G^2}-G^2}\right]\cr
&=& T\left[1-\sqrt{1+\frac{\vec{B}^2-\vec{E}^2}{T}\cosh\gamma-\sinh\gamma\sqrt{\frac{(\vec{B}^2-\vec{E}^2)^2}{T^2}+\frac{4(\vec{B}\cdot
\vec{E})^2}{T^2}}-\frac{(\vec{B}\cdot\vec{E})^2}{T^2}}\right].\cr
&&\label{genLagAns1}
\eea

In the expansion of the above, the $T$-independent terms are definitely correct, since the small field (or $T\rightarrow\infty$)
limit gives the ModMax theory, but the next order terms could in principle not be. However, at least its $\gamma=0$ limit 
is correct, since then we must obtain the BI theory. It is too complicated to check the correctness of the above ansatz directly, however.

Instead, we must use indirect methods to prove it, as we will shortly do. 

Actually, the Lagrangian (\ref{genLagAns1}), corresponding to the Hamiltonian (\ref{genHam}), was also found already in 
\cite{Bandos:2020hgy}, via yet another indirect method.\footnote{A fact that I realized only recently, shortly before finishing this paper.}

We can also calculate back the Hamiltonian of this Lagrangian (i.e., assuming the Lagrangian is correct, find the $\vec{D}, \vec{H}$ and
${\cal H}$ from it), via first obtaining from (\ref{genLagAns1}) that
\bea
\vec{H}=-\frac{\d {\cal L}}{\d \vec{B}}&=&\frac{T\left[\vec{B}\left(\cosh\gamma +\frac{S\sinh\gamma}{\sqrt{S^2+P^2}}\right)
-\vec{E}P\left(\frac{1}{T}+\frac{\sinh\gamma}{\sqrt{S^2+P^2}}\right)\right]}{\sqrt{T^2-2T(S\cosh\gamma +\sinh\gamma \sqrt{S^2+
P^2})-P^2}}\cr
\vec{D}=\frac{\d {\cal L}}{\d \vec{E}}&=&\frac{T\left[\vec{E}\left(\cosh\gamma +\frac{S\sinh\gamma}{\sqrt{S^2+P^2}}\right)
+\vec{B}P\left(\frac{1}{T}+\frac{\sinh\gamma}{\sqrt{S^2+P^2}}\right)\right]}{\sqrt{T^2-2T(S\cosh\gamma +\sinh\gamma \sqrt{S^2+
P^2})-P^2}}\;,\label{backED}
\eea
and then 
\bea
{\cal H}&=&\vec{E}\cdot\vec{D}-{\cal L}=-T\cr
&&+\frac{T \left\{1+\frac{\vec{B}^2}{T}\cosh\gamma+\frac{\sinh \gamma}{T^2\sqrt{\frac{(\vec{B}^2-\vec{E}^2)^2}{T^2}+
4\frac{(\vec{B}\cdot\vec{E})^2}{T^2}}}\left[-\vec{B}^2(\vec{B}^2-\vec{E}^2)-2(\vec{E}\cdot\vec{B})^2\right]\right\}
}{\sqrt{1+\frac{\vec{B}^2-\vec{E}^2}{T}\cosh\gamma-\sinh\gamma\sqrt{\frac{(\vec{B}^2-\vec{E}^2)^2}{T^2}+\frac{4(\vec{B}\cdot
\vec{E})^2}{T^2}}-\frac{(\vec{B}\cdot\vec{E})^2}{T^2}}}\cr
&=&-T
+\frac{T\left[T+\cosh\gamma \vec{B}^2+\frac{\sinh\gamma}{\sqrt{S^2+P^2}}(S\vec{B}^2-P^2)-\frac{P^2}{T}\right]}
{\sqrt{T^2-2T(S\cosh\gamma +\sinh\gamma \sqrt{S^2+P^2})-P^2}}
.\label{backHam}
\eea

The original Hamiltonian (\ref{genHam}) satisfies the duality condition
\be
\vec{E}\cdot\vec{B}=\vec{D}\cdot\vec{H}\;,
\ee
and the condition (needed for conformal invariance of the model, so only in the ModMax $T\rightarrow\infty$ limit)
\be
\vec{E}\times\vec{H}=\vec{D}\times\vec{B}\;,
\ee
while the condition for conformal invariance
\be
\vec{D}\cdot\vec{E}+\vec{B}\cdot\vec{H}=2{\cal H}
\ee
is satisfied only in the $T\rightarrow\infty$ limit. 

Indeed, we can check that, in terms of the Hamiltonian variables, $\vec{B},\vec{D}$, we have 
\bea
\vec{E}=\frac{\d {\cal H}}{\d \vec{D}}&=& \frac{T\left[\cosh \gamma \vec{D} -\frac{\sinh \gamma}{\sqrt{s^2-p^2}}\left(\vec{D}s
-(\vec{D}\vec{B}^2-\vec{B}(\vec{B}\cdot\vec{D}))\right)\right]+\vec{D}\vec{B}^2-\vec{B}(\vec{B}\cdot\vec{D})}{\sqrt{T^2+2T
(s\cosh \gamma -\sinh\gamma \sqrt{s^2-p^2})+p^2}}\cr
\vec{H}=\frac{\d {\cal H}}{\d \vec{B}}&=&  \frac{T\left[\cosh \gamma \vec{B} -\frac{\sinh \gamma}{\sqrt{s^2-p^2}}\left(\vec{B}s
-(\vec{B}\vec{D}^2-\vec{D}(\vec{B}\cdot\vec{D}))\right)\right]+\vec{B}\vec{D}^2-\vec{D}(\vec{B}\cdot\vec{D})}{\sqrt{T^2+2T
(s\cosh \gamma -\sinh\gamma \sqrt{s^2-p^2})+p^2}}\;,\cr
&&
\eea
so that 
\bea
\vec{E}\cdot\vec{B}=\vec{D}\cdot\vec{H}&=&\vec{B}\cdot \vec{D}\frac{\cosh\gamma-\frac{\sinh \gamma s}{\sqrt{s^2-p^2}}}
{\sqrt{T^2+2T(s\cosh \gamma -\sinh\gamma \sqrt{s^2-p^2})+p^2}}\;;\cr
\vec{E}\times\vec{H}&=&\vec{D}\times \vec{B}\;,
\eea
and moreover
\bea
\vec{D}\cdot\vec{E}&=&\frac{T\left[\vec{D}^2\left(\cosh\gamma -\frac{s\sinh \gamma}{\sqrt{s^2-p^2}}\right)+p^2\left(\frac{1}{T}
+\frac{\sinh \gamma}{\sqrt{s^2-p^2}}\right)\right]}{\sqrt{T^2+2T(s\cosh \gamma -\sinh\gamma \sqrt{s^2-p^2})+p^2}}\cr
\vec{B}\cdot\vec{H}&=&\frac{T\left[\vec{B}^2\left(\cosh\gamma -\frac{s\sinh \gamma}{\sqrt{s^2-p^2}}\right)+p^2\left(\frac{1}{T}
+\frac{\sinh \gamma}{\sqrt{s^2-p^2}}\right)\right]}{\sqrt{T^2+2T(s\cosh \gamma -\sinh\gamma \sqrt{s^2-p^2})+p^2}}\;,
\eea
so that
\be
\vec{D}\cdot\vec{E}+\vec{B}\cdot\vec{H}=\frac{2T\left[s\cosh \gamma-\sinh \gamma \sqrt{s^2-p^2}+\frac{p^2}{T}\right]}
{\sqrt{T^2+2T(s\cosh \gamma -\sinh\gamma \sqrt{s^2-p^2})+p^2}}.
\ee

We see that the right-hand side equals $2{\cal H}$ {\em only in the $T\rightarrow\infty$ limit}, i.e. only for the ModMax
theory (expected, since only then the theory is conformal).

But then, the above gives us a way to check the correctness of the ansatz (\ref{genLagAns1}) for the Lagrangian, 
namely that it should satisfy the same relations (and then the uniqueness of the theory, as stated in the original paper, 
ensures that we have the right result). 

From (\ref{backED}), we obtain
\bea
\vec{E}\times\vec{H}=\vec{D}\times \vec{B}&=&\vec{E}\times\vec{B}\frac{T\left(\cosh\gamma+\frac{S\sinh\gamma}{\sqrt{S^2
+P^2}}\right)}{\sqrt{T^2-2T(S\cosh\gamma +\sinh\gamma \sqrt{S^2+P^2})-P^2}}\cr
\vec{D}\cdot\vec{H}&=&\vec{E}\cdot\vec{B}\;,
\eea
and moreover
\bea
\vec{E}\cdot\vec{D}&=&\frac{T\left[\vec{E}^2\left(\cosh\gamma+\frac{S\sinh\gamma}{\sqrt{S^2+P^2}}\right)+P^2\left(\frac{1}{T}
+\frac{\sinh\gamma}{\sqrt{S^2+P^2}}\right)\right]}{\sqrt{T^2-2T(S\cosh\gamma +\sinh\gamma \sqrt{S^2+P^2})-P^2}}\cr
\vec{B}\cdot\vec{H}&=&\frac{T\left[\vec{B}^2\left(\cosh\gamma+\frac{S\sinh\gamma}{\sqrt{S^2+P^2}}\right)-P^2\left(\frac{1}{T}
+\frac{\sinh\gamma}{\sqrt{S^2+P^2}}\right)\right]}{\sqrt{T^2-2T(S\cosh\gamma +\sinh\gamma \sqrt{S^2+P^2})-P^2}}\;,
\eea
so that 
\be
\vec{E}\cdot\vec{D}+\vec{B}\cdot\vec{H}=2\frac{TS\left(\cosh\gamma+\frac{S\sinh\gamma}{\sqrt{S^2+P^2}}\right)}
{\sqrt{T^2-2T(S\cosh\gamma +\sinh\gamma \sqrt{S^2+P^2})-P^2}}\;,
\ee
and the right-hand side is different than (\ref{backHam}), except in the $T\rightarrow\infty$ case, or ModMax case, 
in which it matches, as it should.

\section{Lorentz invariant theory, coupling to a scalar and BIon and catenoid solutions}

We would like to couple the above general model, with Hamiltonian (\ref{genHam}) and Lagrangian (\ref{genLagAns1}) to a 
scalar field, and find its non-magnetic static solutions. As we have previewed, we will find that the same solutions at $\gamma=0$ 
(for the DBI case) are still valid, so we will review the way they are obtained first.

\subsection{DBI Lagrangian and BIon and catenoid solutions} 

In the case of the BI Lagrangian, coupling to scalars 
was easy, and it led to the DBI Lagrangian. Noting that the Lagrangian can be rewritten as
\be
{\cal L}_{\rm BI}(b;\vec{E},\vec{B})=b^2\left[1-\sqrt{\det\left(\eta_{\mu\nu}+\frac{F_{\mu\nu}}{b}\right)}\right]\;,
\ee
the Lorentz invariant coupling is simple, and it leads to
\be
{\cal L}_{DBI}(b;F_{\mu\nu},X)=b^2\left[1-\sqrt{\det\left(\eta_{\mu\nu}+\frac{\d_\mu X\d_\nu X}{b^2}+\frac{F_{\mu\nu}}{b}\right)}\right]\;,
\ee
which moreover, was found to the the action for a D3-brane in string theory moving in flat 5 dimensions, with $X/b$ the transverse 
position of the brane, and $A_\mu$ the open string field on the brane.

Note that the above form of the DBI action is the unique explicitly Lorentz invariant form. Unlike the $X=0$ case, we cannot 
re-express it in terms of the invariants $F$ and $G$, and the invariant $\tilde X=(\d_\mu X)^2/b^2$. 

The most one can say in the DBI case is that, for $\vec{B}=0$, we can rewrite it as \cite{Callan:1997kz}
\be
{\cal L}_{\rm DBI}(b;\vec{E},X)=b^2\left[1-\sqrt{\left(1-\frac{\vec{E}^2}{b^2}\right)\left(1-\frac{(\vec{\nabla}X)^2}{b^2}\right)
+\frac{\left(\vec{E}\cdot \vec{\nabla}X\right)^2}{b^4}-\frac{\dot X^2}{b^2}}\right].
\ee

The momenta for $\vec{A}$ and $X$, respectively, in this case are found to be (in the $A_0=0$ gauge)
\bea
\vec{\Pi}&=&\frac{\d {\cal L}}{\d \vec{E}}=
\frac{\vec{E}\left(1+\frac{(\vec{\nabla} X)^2}{b^2}\right)-\vec{\nabla}X\frac{\vec{E}\cdot\vec{\nabla}X}{b^2}}
{\sqrt{\left(1-\frac{\vec{E}^2}{b^2}\right)\left(1-\frac{(\vec{\nabla}X)^2}{b^2}\right)
+\frac{\left(\vec{E}\cdot \vec{\nabla}X\right)^2}{b^4}-\frac{\dot X^2}{b^2}}}\equiv "\vec{D}(\vec{E},X)"\cr
P&=&\frac{\d {\cal L}}{\d \dot X}=\frac{\dot X}{\sqrt{\left(1-\frac{\vec{E}^2}{b^2}\right)\left(1-\frac{(\vec{\nabla}X)^2}{b^2}\right)
+\frac{\left(\vec{E}\cdot \vec{\nabla}X\right)^2}{b^4}-\frac{\dot X^2}{b^2}}}\;,
\eea
(we see that $\vec{\Pi}$ is a generalization, in the presence of $X$, of $\vec{D}$)
so the Hamiltonian becomes
\be
{\cal H}_{\rm DBI}(b;\vec{\Pi},P,\vec{\nabla}X)=
b^2\left[\sqrt{\left(1+\frac{(\vec{\nabla}X)^2}{b^2}\right)\left(1+\frac{P^2}{b^2}\right)+\frac{\vec{\Pi}^2}{b^2}+\frac{(\vec{\Pi}\cdot \vec{\nabla}
X)^2}{b^4}}-1\right]\;,
\ee
subject to the Gauss constraint $\vec{\nabla}\cdot \vec{\Pi}=0$, due to the absence of the $A_0$ canonical momentum, as usual.

The static equation for $X$, obtained by varying the above Hamiltonian with respect to $X$, 
after putting $P=0$ and $\vec{\nabla}\cdot \vec{\Pi}=0$, is 
\bea
&&\vec{\nabla}\cdot \left[\frac{\vec{\nabla} X+\vec{\Pi}\frac{\vec{\nabla}X \cdot \vec{\Pi}}{b^2}}
{\sqrt{1+\frac{(\vec{\nabla}X)^2}{b^2}+\frac{\vec{\Pi}^2}{b^2}+\frac{(\vec{\Pi}\cdot \vec{\nabla}X)^2}{b^4}}}\right]\cr
&=& \vec{\nabla}\cdot \left[\frac{\vec{\nabla}X \left(1-\frac{\vec{E}^2}{b^2}\right)+\vec{E}\frac{\vec{E}\cdot\vec{\nabla}X}{b^2}}
{\sqrt{\left(1-\frac{\vec{E}^2}{b^2}\right)\left(1+\frac{(\vec{\nabla}X)^2}{b^2}\right)+\frac{(\vec{E}\cdot\vec{\nabla}X)^2}{b^4}}}\right]=0.
\label{DBIeom}
\eea

Static solutions without magnetic field need only solve the above equation and the constraint $\vec{\nabla}\cdot\vec{\Pi}=0$. 
We have (see \cite{Gibbons:1997xz,Gibbons:2001gy} for a review of these solutions):

-1) The BPS (extremal) BIon, corresponding in the string picture to a fundamental string attached to the D-brane, sourced by an electric 
field charge on the D-brane. This is obtained by putting (writing the sum of extremal BIons, which is also a solution, due to the BPS 
nature of it)
\bea
&&\vec{\Pi}=\vec{\nabla}X=\vec{\nabla} \Lambda\;,\;\;
\Lambda=\sum_i \frac{q_i}{|\vec{r}-\vec{r}_i|}\Rightarrow\cr
&&X=\sum_i\frac{q_i}{|\vec{r}-\vec{r}_i|}\;,\;\; \vec{E}=\vec{\nabla}X.
\eea

It is easy to see that it is a solution to both the constraint and (\ref{DBIeom}). Its energy linearizes, and is 
\be
E=\int d^3x (\vec{\nabla} \Lambda)^2\;,
\ee
and gives rise to a finite energy.

-2) The usual BIon solution of Born and Infeld, with $X=0$, and $\vec{E}=-\vec{\nabla}\phi$, and 
\be
\phi=C\int_r^\infty \frac{dx}{\sqrt{C^2b^2+x^4}}\;,
\ee
which becomes $\phi\rightarrow \frac{C}{r}$ at $r\rightarrow \infty$. This solution has also a finite energy, which
was identified by Born and Infeld with the energy of a non-divergent electron. 

-3) The "catenoid" solution, describing one half of a D-brane-anti-D-brane bridge, with $\vec{E}=0$ and 
\be
X=\tilde C\int_r^\infty \frac{dx}{\sqrt{x^4-b^2\tilde C^2}}\;,
\ee
and also becomes $X\rightarrow \frac{\tilde C}{r}$ at $r\rightarrow\infty$.
This has a "horizon" (where we glue to another solution) at $r_0=\sqrt{b\tilde C}$, where $\vec{F}\equiv \vec{\nabla}X$ diverges, 
though $X$ doesn't, so the solution makes sense only for $r>r_0$.

The most general static spherically symmetric solution is obtained by writing in this case the constraint $\vec{\nabla}\cdot\vec{\Pi}=0$ 
and the equation of motion (\ref{DBIeom}), which become
\be
\frac{d}{dr}\left[\frac{r^2\phi'}{\sqrt{1-\frac{\phi'^2}{b^2}+\frac{X'^2}{b^2}}}\right]=0\;,\;\;\;\;
\frac{d}{dr}\left[\frac{r^2X'}{\sqrt{1-\frac{\phi'^2}{b^2}+\frac{X'^2}{b^2}}}\right]=0\;,
\ee
which can be integrated with integration constants $C$ and $\tilde C$, respectively, leading to the most general solution
\be
\phi=C\int_r^\infty \frac{dx}{\sqrt{x^4+b^2(C^2-\tilde C^2)}}\;,\;\;
X=\tilde C\int_r^\infty \frac{dx}{\sqrt{x^4+b^2(C^2-\tilde C^2)}}.
\ee

We see that $C^2<\tilde C^2$ gives a catenoid-like solution, with $C=0$ giving the catenoid, $C^2>\tilde C^2$ gives a 
BIon-like solution, with $\tilde C=0$ being the BIon, and $C=\tilde C$ gives the BPS BIon solution corresponding to the 
fundamental string ending on the D-brane. 

\subsection{Lorentz invariant theory and coupling to a scalar}

In order to couple to a scalar, we must write a Lorentz invariant form of the action. But, as we saw in the DBI case, that is only 
possible if we have it in a determinant-type form. Otherwise, if we try to rewrite it in terms of the Lorentz invariants $F$ and $G$, 
that fails once the scalar is introduced. Then the thing we expect to have  inside the square root would be of the form 
\be
\det\left(\eta_{\mu\nu}+\a (F,\tilde F)\frac{F_{\mu\nu}}{b}+\b(F,\tilde F)\frac{\tilde F_{\mu\nu}}{b}\right)=
\det\left(\eta_{\mu\nu}+\a(F,\tilde F) \frac{F_{\mu\nu}}{b}+\frac{\b(F,\tilde F)}
{2}\epsilon_{\mu\nu\rho\sigma}\frac{F^{\rho\sigma}}{b}\right)
\;,
\ee
i.e., starting with the standard BI action, replace
\be
F_{\mu\nu}\rightarrow\a(F,\tilde F) F_{\mu\nu}+\b(F,\tilde F)\tilde F_{\mu\nu}\;,
\ee
which gives rise to the change (in terms of $A=F_{\mu\nu}F^{\mu\nu}$ and $B=F_{\mu\nu}\tilde F^{\mu\nu}$, or more 
precisely $F=A/(2b^2)$ and $G=-B/(4b^2)$)
\be
F\rightarrow (\a^2-\b^2)F-4\a\b G\;,\;\;\;
G\rightarrow (\a^2-\b^2)G+\a\b F.
\ee

But given that the determinant in the BI action gives 
\be
1+F-G^2\;,
\ee
and it is supposed to (we want it to) get modified into
\be
1+F\cosh\gamma-\sinh\gamma \sqrt{F^2+4G^2}-G^2\;,
\ee
we obtain two equations for two unknowns, $\a$ and $\b$, 
\bea
(\a^2-\b^2)F-4\a\b G&=& \cosh\gamma F-\sinh\gamma \sqrt{F^2+4G^2}\cr
(\a^2-\b^2)G+\a\b F&=&\pm G.
\eea

To solve them, define
\be
a\equiv \a^2-\b^2\;,\;\; b\equiv \a\b\;,
\ee
and first write $a=1-bF/G$, then find 
\bea
\frac{(\a+\b)^2-(\a-\b)^2}{4}=\a\b&=&b=\frac{FG}{F^2+4G^2}(1-\cosh\gamma)+\sinh\gamma \frac{G}{\sqrt{F^2+4G^2}}\cr
(\a+\b)(\a-\b)=\a^2-\b^2&=&a=1-\frac{F^2}{F^2+4G^2}(1-\cosh\gamma)-\sinh\gamma \frac{F}{\sqrt{F^2+4G^2}}.\cr
&&
\eea

We want to avoid singularities in both the $G=0$ case (for instance, for $\vec{B}=0$), and $F=0$ case (null, for instance electromagnetic 
wave). 

Indeed, for $G=0$ (and $F\neq 0$), we have $a=e^{-\gamma}, b=0$, so 
\be
\a=e^{-2\gamma}\;,\;\;\b=0.
\ee

Also, for $F=0$ (and $G\neq 0$), we have $b=\frac{1}{2}\sinh\gamma, a=1$, giving 
\be
\a=\cosh\frac{\gamma}{2}\;,\;\;\;
\b=\sinh\frac{\gamma}{2}.
\ee

The only potential singularity is for the case $F=G=0$, just like for the limit ModMax theory. 

Finally, coupling to the scalar is done in the usual way for the DBI action, assuming that that the scalar $X$ part is not 
modified, so 
\be
{\cal L}(X,F_{\mu\nu})=b^2\left[1-\sqrt{-\det\left(\eta_{\mu\nu}+\frac{\d_\mu X\d_\nu X}{b^2}+\a(F,G)\frac{F_{\mu\nu}}{b}+\b(F,G)
\frac{\tilde F_{\mu\nu}}{b}\right)}\right].
\ee

Since inside the square root we obtain $1+(\d_\mu X)^2/b^2+F$ terms, when taking the large $T=b^2$ limit, we obtain 
only a free scalar action, $-\frac{1}{2}(\d_\mu X)^2$, added to the ModMax theory, so it is not so interesting. 

It is much more interesting (meaning, we have interactions to between the scalar and gauge fields), if we keep $T$ finite, so
that is what we will consider next.

\subsection{Static solutions with no magnetic field}

Since we are interested in solutions at $\vec{B}=0$, consider
$\vec{B}=0$ in the Lagrangian (\ref{genLagAns1}), giving
\be
{\cal L}=T\left[1-\sqrt{1-\frac{e^\gamma}{T}\vec{E}^2}\right]\;,
\ee
which by the way is
consistent with the fact that when $\vec{B}=0$, the Hamiltonian (\ref{genHam}) becomes (then $p=0,2s=\vec{D}^2$)
\be
{\cal H}=T\left[\sqrt{1+\frac{\vec{D}^2}{Te^\gamma}}-1\right]\;.\label{HamnoB}
\ee

Then (since up to the $e^\gamma$ term, this is the same as in the DBI case) it follows that we can use the same logic as in the DBI
case, and say that, in the $A_0=0$ gauge, we have the equation of motion reducing to 
\be
\vec{\nabla}\cdot\vec{D}=0\;,
\ee
or rather delta function sources on the right-hand side, which is solved as usual by by 
\be
\vec{D}=\vec{\nabla}\frac{q}{|\vec{x}-\vec{x}_0|}\;,
\ee
except now there are some factors of $e^\gamma$ in the relation between $\vec{E}$ and $\vec{D}$, 
\be
\vec{E}=\frac{\d {\cal H}}{\d \vec{D}}=\frac{\vec{D}}{e^\gamma\sqrt{1+\frac{\vec{D}^2}{Te^\gamma}}}=e^{-\gamma}
\frac{\vec{D}}{\sqrt{1+\frac{\vec{D}^2}{\tilde T}}}.
\ee

We can then add $X$ directly in the Hamiltonian at $\vec{B}=0$, and obtain ($\tilde T=e^\gamma T$)
\be
{\cal H}(\vec{\Pi},P,X)=e^{-\gamma}\tilde T\left[\sqrt{\left(1+\frac{(\vec{\nabla}X)^2}{\tilde T}\right)\left(1+\frac{P^2}{\tilde T}\right)
+\frac{\vec{\Pi}^2}{\tilde T}+\frac{(\vec{\Pi}\cdot\vec{\nabla}X)^2}{\tilde T^2}}-1\right].
\ee

Then the static solutions at $\vec{B}=0$ are the same ones as for the usual DBI case, just rescaled by $e^{-\gamma}$, 
namely ($\tilde b^2=\tilde T$) in the most general case
\be
\phi=e^{-\gamma}C\int_r^\infty \frac{dx}{\sqrt{x^4+\tilde b^2(C^2-\tilde C^2)}}\;,\;\;
X=e^{-\gamma}\tilde C\int_r^\infty \frac{dx}{\sqrt{x^4+\tilde b^2(C^2-\tilde C^2)}}\;,
\ee
$\vec{E}=-\vec{\nabla}\phi$, with $C=0$ giving the catenoid, $\tilde C=0$ the BIon, and $C=\tilde C$ the extremal BIon.

\subsection{Knotted solutions to precursor of ModMax}

Maxwell's equations in vacuum have solutions with nonzero helicities 
\bea
H_{mm}&=&\int d^3x \epsilon^{ijk}A_i\d_j A_k\;,\cr
H_{ee}&=&\int d^3x \epsilon^{ijk}C_i\d_j C_k\;, \cr
H_{em}&=&\int d^3x \epsilon^{ijk}C_i \d_j A_k\;, \cr
H_{me}&=&\int d^3x \epsilon^{ijk}A_i\d_j C_k\;,
\eea
where $\vec{B}=\vec{\nabla}\times \vec{A}$ and $\vec{E}=\vec{\nabla}\times \vec{C}$, with nonzero Hopf index, and 
perhaps knotted (with nonzero linking numbers). The basic Hopfion 
(time-dependent) solution was found in Bateman's formalism \cite{Bateman:1915}, via 
the ansatz
\be
\vec{F}\equiv \vec{E}+i\vec{B}=\vec{\nabla}\a \times\vec{\nabla}\b\;,
\ee
where $\a$ and $\b$ are specific complex functions of spacetime variables,
\be
\a=\frac{A-1+iz}{A+it}\;,\;\;
\b=\frac{x-iy}{A+it}\;,\;\;
A=\frac{1}{2}(x^2+y^2+z^2-t^2+1)\;,
\ee
and $(p,q)$-knotted solutions are obtained by the replacement $\a\rightarrow\a^p$, $\b\rightarrow\b^q$, or by using acting on a 
plane wave (unknotted) solution with conformal symmetry transformations with {\em complex} instead of real parameters 
\cite{Hoyos:2015bxa}. The Hopfion and knotted Hopfion solution have $F=G=0$ and, since 
\bea
\d_t H_{mm}&=&-2\int d^3x \vec{E}\cdot\vec{B}=\d_t H_{ee}\cr
\d_t H_{me}&=&-\int d^3x (\vec{E}^2-\vec{B}^2)=-\d_t H_{em}\;,
\eea
all the helicities are conserved on the solution. 

More general solutions were found by Ra\~nada \cite{Ranada:1989wc,ranada1990knotted}. One set of solutions 
have $G=0$ ($\vec{E}\cdot\vec{B}=0$), but $F$ arbitrary ($\vec{E}^2-\vec{B}^2\neq 0$), so 
have only $H_{ee}$ and $H_{mm}$ conserved on the solution, and are given by 
\bea
\vec{B}(\vec{r},t)&=& g(\phi,\bar\phi)\vec{\nabla}\phi\times \vec{\nabla}\bar\phi\cr
\vec{E}(\vec{r},t)&=&g'(\theta,\bar\theta)\vec{\nabla}\theta\times\vec{\nabla}\bar\theta\cr
g(\phi,\bar\phi)&=&\frac{\sqrt{a}}{2\pi i}\frac{1}{(1+\bar\phi\phi)^2}\;,\;\;
g'(\theta,\bar\theta)=\frac{\sqrt{a}}{2\pi i}\frac{1}{(1+\bar\theta\theta)^2}\;,\label{Ranada}
\eea
satisfy $\d_{[\mu}F_{\nu\rho]}=0$, and can be rewritten in terms of real functions $p,q,u,v$ (defined only on 
patches in whole space) as 
\bea
\vec{B}&=&\vec{\nabla}p\times \vec{\nabla}q\;,\;\;
p=\frac{1}{1+|\phi|^2}\;,\;\; q=\frac{{\rm arg}(\phi)}{2\pi}\;,\cr
\vec{E}&=& \vec{\nabla}u\times \vec{\nabla v}\;,\;\;
u=\frac{1}{1+|\theta|^2}\;,\;\; v=\frac{{\rm arg}(\theta)}{2\pi}.
\eea

There are also another set of solutions, with both $F\neq 0$ and $G\neq 0$ ($\vec{E}^2-\vec{B}^2\neq 0$ and $\vec{E}\cdot\vec{B}\neq 0$)
found implicitly (no explicit form is given, only a condition on the form at time $t=0$) from the Ra\~nada construction in \cite{Arrayas:2015aa}, 
but we will not be interested in them here. In \cite{Dassy:2021ulu} there was a direct construction for deformed Ra\~nada type solutions 
for the ModMax theory, that reduce to the Ra\~nada solutions at $\gamma=0$. 

However, in \cite{Alves:2017ggb,Alves:2017zjt}, the following observation was made. Since the equations of motion of the nonlinear 
electromagnetic theory written in terms of $F$ and $G$ can be put into the form of the Maxwell equations in a medium, in terms of 
$\vec{E},\vec{B},\vec{D},\vec{H}$, as in (\ref{Mxmedium}), all we need to do is ensure that 
\be
\vec{H}=-\frac{\d {\cal L}}{\d \vec{B}}=\vec{B}+{\cal O}(F,G)\;,\;\;\;
\vec{D}=\frac{\d {\cal L}}{\d \vec{E}}=\vec{E}+{\cal O}(F,G)\;,
\ee
for the same Bateman solutions with $F=G=0$ to be still valid. That is satisfied for a Lagrangian that admits a Taylor expansion
around $F=G=0$ and reduces to Maxwell in the small field limit, i.e.,
\be
{\cal L}(F,G)=b^2\left[-\frac{F}{2}+\sum_{n\geq 2}\sum_{m\geq 0}c_{n,m} F^n G^m\right].
\ee

In particular, this applied to the BI theory, for which $b^2[1-\sqrt{1+F-G^2}]$ can be expanded as above. 

In the case of the ModMax theory, however, the limit $(F,G)\rightarrow (0,0)$ is singular because of the square root, so the above 
seems not to hold. However, in this case of the ModMax theory, we observe that we have 
\bea
{\cal L}(F,G)&=&-\frac{Fb^2}{2}\cosh \gamma +\frac{\sinh\gamma}{2} f(F^2,G^2)\;,\;\;
{\cal L}(F,G=0)=-\frac{Fb^2}{2}e^\gamma \Rightarrow\cr
\vec{H}(G=0)&=&e^\gamma \vec{B}\;,\;\; \vec{D}(G=0)=e^\gamma\vec{E}.\label{gammaEB}
\eea

But the factors of $e^\gamma$ in $\vec{H}$ and $\vec{D}$ can be reabsorbed in a trivial redefinition, in order to obtain 
the same equations of motion as in vacuum, since classically a multiplicative constant in front of the Lagrangian is irrelevant. 

That means that the Ra\~nada solutions (\ref{Ranada}) are still valid for the ModMax theory. 
When going from ModMax to its precursor (\ref{genLagAns1}), the same logic as  \cite{Alves:2017ggb,Alves:2017zjt}, described 
above, for going from Maxwell to BI, still applies. Namely, the expanded Lagrangian is now
\be
{\cal L}(F,G)={\rm ModMax}(F,G)+\sum_{n\geq 2}\sum_{m\geq 0}c_{n,m}[{\rm ModMax}(F,G)]^n G^{2m}\;,
\ee
and so again, it leads to (\ref{gammaEB}), hence the Ra\~nada solutions with $G=0, F\neq 0$ (\ref{Ranada}) are still valid.

\section{Conclusions}

In this paper, I have shown how to couple the precursor to the ModMax model to scalars, by writing it in a square root of determinant
form, similar to the DBI action. For static solutions in the absence of magnetic fields, the action is the same as in the DBI case, 
which led to the same solutions being still valid: the supersymmetric (BPS) BIon, the original BIon, and the catenoid, as well as 
all the general solutions in between. 
I have also shown that the Ra\~nada knotted solutions with $F\neq 0, G=0$ are still solutions of both the ModMax theory and 
its precursor (and, of course, the full action with scalars as well, of which that is a consistent truncation, as the action contains
only terms quadratic in scalars).

An important question remains: can one still understand this precursor to the ModMax model somehow in string theory, as a 
modification of the original DBI brane action? It is not clear, since the form we wrote for the action had some nontrivial 
scalar coefficients $\a$ and $\b$ depending on the Lorentz invariants $F$ and $G$ (or $A$ and $B$) made from the electromagnetic 
field, and it is not clear how under what conditions these could be obtained.

\section*{Acknowledgements}

I would like to thank Cobi Sonnenschein for discussions.
My work is supported in part by CNPq grant 301491/2019-4 and FAPESP grants 2019/21281-4 
and 2019/13231-7. I would also like to thank the ICTP-SAIFR for their support through FAPESP grant 2016/01343-7.

\appendix


\section{Quick review of BI electrodynamics}

The Born-Infeld theory of electromagnetism \cite{Born:1934gh} is obtained for $\gamma=0$ in (\ref{genHam}), so it is worth 
reviewing what happens in this case, in terms of the formalism above, in order to generalize it.

The Lagrangian is 
\be
{\cal L}(b;\vec{E},\vec{B})=b^2\left[1-\sqrt{1+F-G^2}\right]\;,\label{LagDBI}
\ee
where 
\be
F=\frac{1}{b^2}(\vec{B}^2-\vec{E}^2)=\frac{1}{2b^2}F_{\mu\nu}F^{\mu\nu}\;,\;\;
G=\frac{1}{b^2}\vec{E}\cdot\vec{B}=-\frac{1}{4b^2}F_{\mu\nu}\tilde F^{\mu\nu}\;,
\ee
with $\tilde F^{\mu\nu}=\frac{1}{2}\epsilon^{\mu\nu\rho\sigma}F_{\rho\sigma}$.

The constitutive relations become 
\be
\vec{H}=-\frac{\d {\cal L}}{\d \vec{B}}=\frac{\vec{B}-G\vec{E}}{\sqrt{1+F-G^2}}\;,\;\;
\vec{D}=\frac{\d {\cal L}}{\d \vec{E}}=\frac{\vec{E}+G\vec{B}}{\sqrt{1+F-G^2}}.
\ee

Then the Hamiltonian is 
\be
{\cal H}=\vec{E}\vec{D}-{\cal L}=b^2\left[\frac{1+\frac{\vec{B}^2}{b^2}}{\sqrt{1+\frac{\vec{B}^2-\vec{E}^2}{b^2}-\left(\frac{\vec{B}\cdot
\vec{E}}{b^2}\right)^2}}-1\right]\;,
\ee
but it needs to be re-expressed in terms of $\vec{D},\vec{B}$, where $\vec{D}$ is now explicitly
\be
\vec{D}=\frac{\vec{E}+\vec{B}\left(\frac{\vec{E}\cdot\vec{B}}{b^2}\right)}{\sqrt{1+\frac{\vec{B}^2-\vec{E}^2}{b^2}-\left(\frac{\vec{B}\cdot
\vec{E}}{b^2}\right)^2}}\;,
\ee
and we can thus calculate
\bea
2s&= & \vec{D}^2+\vec{B}^2=\frac{\vec{E}^2+\vec{B}^2\left(1+\frac{\vec{B}^2-\vec{E}^2}{b^2}\right)+2\frac{(\vec{E}\cdot\vec{B})^2}{b^2}}
{1+\frac{\vec{B}^2-\vec{E}^2}{b^2}-\left(\frac{\vec{B}\cdot\vec{E}}{b^2}\right)^2}\cr
p^2&=&\vec{D}^2\vec{B}^2-(\vec{B}\cdot\vec{D})^2=
\frac{\vec{E}^2\vec{B}^2-\frac{(\vec{E}\cdot\vec{B})^2}{b^2}}{1+\frac{\vec{B}^2-\vec{E}^2}{b^2}
-\left(\frac{\vec{B}\cdot\vec{E}}{b^2}\right)^2}\;,
\eea
which allows us to check that
\be
{\cal H}(b;\vec{D},\vec{B})=b^2\left[\sqrt{1+\frac{2s}{b^2}+\frac{p^2}{b^4}}-1\right]=b^2\left[\sqrt{1+\frac{\vec{D}^2+\vec{B}^2}{b^2}
+\frac{\vec{D}^2\vec{B}^2-(\vec{D}\cdot\vec{B})^2}{b^4}}-1\right].
\ee

We now note, by comparing with the $\gamma=0$ limit of (\ref{genHam}), that we have $T=b^2$.

We can also, reversely, calculate
\be
\vec{E}=\frac{\d {\cal H}(b;\vec{B},\vec{D})}{\d \vec{D}}=\frac{\vec{D}+\frac{\vec{D}\vec{B}^2-\vec{B}(\vec{D}\cdot\vec{B})}{b^2}}{
\sqrt{1+\frac{\vec{D}^2+\vec{B}^2}{b^2}+\frac{\vec{D}^2\vec{B}^2-(\vec{B}\cdot\vec{D})^2}{b^4}}}\;,
\ee
and then 
\be
{\cal L}(b;\vec{E},\vec{B})=\vec{D}\cdot\vec{E}-{\cal H}(b;\vec{D},\vec{B})=b^2\left[1-\frac{1+\frac{\vec{B}^2}{b^2}}
{\sqrt{1+\frac{\vec{D}^2+\vec{B}^2}{b^2}+\frac{\vec{D}^2\vec{B}^2-(\vec{B}\cdot\vec{D})^2}{b^4}}}\right]\;,
\ee
and then we can check that we can rewrite it in terms of $\vec{E},\vec{D}$ as in (\ref{LagDBI}).

\bibliography{ModMaxpaper}
\bibliographystyle{utphys}

\end{document}